# Tunable positions of Weyl nodes via magnetism and pressure in the ferromagnetic Weyl semimetal CeAlSi

Erjian Cheng[1,*,☯], Limin Yan[2,3,*], Xianbiao Shi[4,5,*], Rui Lou[6,7,8,*,ǂ], Alexander Fedorov[1,7,8], Mahdi Behnami[1], Jian Yuan[9], Yuanji Xu[10],Yang Xu[11], Wei Xia[9], Nikolai Pavlovskii[12], Darren C. Peets[12], Weiwei Zhao[4,5], Yimin Wan[13], Yanfeng Guo[9], Shiyan Li[13,14,15], Wenge Yang[2,¶] and Bernd Büchner[1,16,§]

[1]*Leibniz Institute for Solid State and Materials Research (IFW-Dresden), 01069 Dresden, Germany*

[2]*Center for High Pressure Science and Technology Advanced Research, 201203 Shanghai, China*

[3]*State Key Laboratory of Superhard Materials, Department of Physics, Jilin University, 130012 Changchun, China*

[4]*State Key Laboratory of Advanced Welding & Joining, Harbin Institute of Technology, 150001 Harbin, China*

[5]*Flexible Printed Electronics Technology Center, Harbin Institute of Technology (Shenzhen), 518055 Shenzhen, China*

[6]*School of Physical Science and Technology, Lanzhou University, 730000 Lanzhou, China*

[7]*Helmholtz-Zentrum Berlin für Materialien und Energie, Albert-Einstein-Straße 15, 12489 Berlin, Germany*

[8]*Joint Laboratory "Functional Quantum Materials" at BESSY II, 12489 Berlin, Germany*

[9]*School of Physical Science and Technology, ShanghaiTech University, 200031 Shanghai, China*

[10]*Institute for Applied Physics, University of Science and Technology Beijing, 100083 Beijing, China*

[11]*Key Laboratory of Polar Materials and Devices (MOE), School of Physics and Electronic Science, East China Normal University, 200241 Shanghai, China*

[12]*Institute of Solid State and Materials Physics, Technische Universität Dresden, 01069 Dresden, Germany*

[13]*State Key Laboratory of Surface Physics, and Department of Physics, Fudan University, 200438 Shanghai, China*

[14]*Collaborative Innovation Center of Advanced Microstructures, 210093 Nanjing, China*

[15]*Shanghai Research Center for Quantum Sciences, 201315 Shanghai, China*

[16]*Institute of Solid State and Materials Physics and Würzburg-Dresden Cluster of Excellence—ct.qmat, Technische Universität Dresden, 01062 Dresden, Germany*

## Abstract

**The noncentrosymmetric ferromagnetic Weyl semimetal CeAlSi with simultaneous space-inversion (SI) and time-reversal (TR) symmetry breaking provides a unique platform for the exploration of novel topological states. Here, by employing electrical and thermoelectrical transport, angle-resolved photoemission spectroscopy (ARPES), high-pressure techniques, and band calculations, we demonstrate that magnetism and pressure can serve as efficient parameters to tune the positions of Weyl nodes in CeAlSi. At ambient pressure, an anomalous Hall effect (AHE) and an anomalous Nernst effect (ANE) arise in the paramagnetic state, and then are enhanced when temperature approaches the ferromagnetic ordering temperature, evidencing magnetism facilitates the AHE/ANE. Such an enhancement of AHE/ANE can be ascribed to the tuning of the positions of Weyl nodes via magnetism. The ARPES measurements reveal that the ferromagnetism serves as a pivotal knob to tune the band structure of CeAlSi both in the bulk and on the surface. Such magnetism-tunable electronic structure has hitherto not been reported in other magnetic *R*AlPn (*R* = rare earth elements, *Pn* = Si, Ge) siblings, suggesting the great potential of controlling Weyl node positions in CeAlSi. Under pressure, an enhancement and a sign change of AHE are discovered. Based on band calculations, the evolution of AHE may root in the tuning of Weyl nodes via pressure. Moreover, multiple pressure-induced phase transitions are uncovered. These findings indicate that CeAlSi provides a unique and tunable platform for exploring exotic topological physics and electron correlations, as well as catering to an array of potential applications, such as spintronics and thermoelectrics.**

## Introduction

Over the last decade, the advancement of theoretical predictions and experimental validations in topological semimetals has hastened the progress of research on topological states of matter and topotronics[1–4]. In topological materials, particularly significant efforts have been devoted to searching for and characterizing novel topological states due to their exotic properties, such as the presence of low-energy excitations, extremely large and unsaturated magnetoresistance (MR), topological surface states, Fermi arcs, chiral anomaly[1–4]. The addition of magnetic elements in magnetic topological materials (MTMs) breaks the time-reversal (TR) symmetry, leading to various intriguing phenomena[5–9], such as intrinsic anomalous Hall/Nernst[1–4,10] and topological Hall/Nernst effects[11–17], and topological magnetic textures (for example, skyrmions[11–18], hedgehogs[19,20], merons[21], magnetic bubbles[13], hopfions[22]). The interplay between magnetism and topology remains mysterious but is expected to be promising for the realization of novel topological states. Hitherto research on this interplay is limited to a few cases, for example, a magnetic-field-induced ideal type-II Weyl state in $Mn(Bi_{1-x}Sb_x)_2Te_4$ [23,24], a magnetic-exchange-induced Weyl state in $EuCd_2Sb_2$ [25], a spin-fluctuation-induced Weyl semimetal state in $EuCd_2As_2$ [17,26], a magnetism-induced topological transition in $EuAs_3$ [27], and magnetization-tunable Weyl states in $EuB_6$ [28], etc. To exploit more novel phenomena and elaborate the relationship, more systems are called for.

Recently, the noncentrosymmetric *R*Al*Pn* series were proposed and demonstrated to be Weyl semimetals[21,29–49], which host low-energy excitations, namely Weyl fermions described by the Weyl equation with $2 \times 2$ complex Pauli matrices[1–4]. Weyl fermions arise in the vicinity of doubly degenerate electronic band crossing point (the Weyl node), and a pair of Weyl nodes possess opposite chirality[1–4]. In general, to attain Weyl states, space-inversion (SI) or TR symmetry should be broken, and there are a few cases in which SI and TR symmetries are simultaneously broken[1–4]. For the nonmagnetic LaAl*Pn*, SI symmetry is naturally broken, and the system hosts two types of Weyl states (type-I and type-II)[31,32]. Moreover, a spin Hall angle that is comparable to $MTe_2$ ($M$ = W, Mo) has been predicted in LaAl*Pn*[50–54]. More intriguingly, pressure-induced superconductivity and robust topology against pressure up to 80.4 GPa have been uncovered in LaAl*Pn*[55]. In contrast to LaAl*Pn*, both SI and TR symmetries are broken in the magnetic siblings, i.e., *R*Al*Pn* (*R* = Ce, Pr, Nd, Sm, *Pn* = Si, Ge), rendering them rare cases for studying novel topological properties with the simultaneous breaking of SI and TR symmetries[21,29,36,39,49,56].

CeAlSi is a ferromagnetic Weyl semimetal with noncollinear magnetic ordering, and electrical transport measurements revealed an anisotropic AHE, a loop-shaped Hall effect (LHE) in the ferromagnetic state, and a nontrivial Berry phase[41]. Previous ARPES experiments in the paramagnetic phase unveiled surface Fermi arcs and linearly dispersing conical features that correspond to the Weyl cones, further demonstrating the existence of Weyl fermions[42]. The flat band stemming from Ce 4*f* electrons was also detected near Fermi level ($E_F$), indicating that electron correlations may also play a role[42]. Scanning superconducting quantum interference device (sSQUID) and magneto-optical Kerr effect (MOKE) microscopy on CeAlSi found the presence of nontrivial chiral domain walls that contributed to the topological properties[43–45]. According to DFT calculations, the Weyl nodes in CeAlSi arise from the SI symmetry breaking, and the TR symmetry breaking with the inclusion of magnetism just shifts the positions of Weyl nodes in the Brillouin zone (BZ) as the ferromagnetism acts as a simple Zeeman coupling[30,41], which lacks experimental verification. Upon compressing CeAlSi up to ~ 3 GPa, it was proposed that the electronic structure and magnetic structure of CeAlSi remain nearly unchanged, but the AHE and LHE are suppressed[45]. Nevertheless, the evolution of band structure and topology under higher pressure has not been elaborated[40,45]. Since ferromagnetism is proposed to serve as an efficient parameter to tune the positions of Weyl nodes, therefore how and to what extent the Weyl nodes in CeAlSi evolve with pressure remains elusive.

In this work, by resorting to electrical and thermoelectrical transport, ARPES, high-pressure techniques, and band calculations, we systematically study the band structure and topological properties of CeAlSi. At ambient pressure, both ANE and AHE in CeAlSi are unveiled. They arise in the paramagnetic state, and then are enhanced when temperature approaches the ferromagnetic transition temperature ($T_C$), indicative of the interaction of magnetism and topology. The anomalous Hall conductivity ($\sigma_{xy}^A$) and the anomalous Nernst conductivity ($\alpha_{xy}^A$) follow the Mott relation, and the latter is the derivative of the former at $E_F$[10,57]. When $E_F$ crosses the Weyl points, $\sigma_{xy}^A$ reaches a maximum, while $\alpha_{xy}^A$ peaks when $E_F$ shifts away from the Weyl points[57]. Since the TR symmetry breaking due to the inclusion of ferromagnetism in CeAlSi does not change the classification and the quantity of Weyl nodes but just shifts their positions, the enhancement of AHE/ANE may arise from the shift of the positions of Weyl nodes. The magnetic tunability of both the bulk and surface band structure of CeAlSi is unambiguously evidenced by our ARPES experiments, further illuminating that the Weyl node positions can be adjusted by magnetism. Under pressure, an enhancement and a sign change of AHE take place. Based on band calculations, we found that pressure has similar effect on Weyl nodes as magnetism. In addition, multiple pressure-induced phase transitions are discovered, i.e., a pressure-induced Lifshitz transition at ~ 10 GPa, a magnetic transition from the ferromagnetic state to a paramagnetic state beyond ~ 20 GPa, and a structural phase transition at ~ 40 GPa. These results suggest that CeAlSi provides a unique and tunable platform to study novel topological states, the interplay between magnetism and topology, and topological properties with electron correlation effect.

## Results

### Anomalous magneto-transverse transport in CeAlSi

CeAlSi crystallizes in the tetragonal structure with the space group $I4_1md$ (No. 109), as shown in Fig. 1(a)[32,41]. High-quality single crystals of CeAlSi are synthesized through a flux method[32,41]. The largest natural surface is the *ab* plane [Supplementary Fig. 2(a)]. CeAlSi possesses in-plane (the *ab* plane) noncollinear ferromagnetic ordering below ~ 10 K defined according to magnetization, which is also evident in resistivity [Fig. 1(c)][41]. Figure 1(b) displays the calculated electronic band structure and associated Berry curvature. Figure 1(d) shows the Hall resistivity of CeAlSi at different temperatures. There is a turning point at ~ 2.5 T, above which the Hall resistivity profile with a positive slope displays a linear dependence. The turning point in Hall resistivity persists up to ~ 10 K, and then broadens and shifts to higher fields. Above ~ 100 K, the Hall resistivity displays linear behavior. Figure 1(e) shows the magnetization at different temperatures. The magnetization above 15 K displays a linear dependence, evidencing that CeAlSi is in the paramagnetic state. When temperature decreases below 15 K, the system approaches the regime of magnetic fluctuations, and nonlinear components start to contribute. To obtain the anomalous contributions in Hall resistivity, we subtract the linear background by adopting the expression, $\rho_{yx} = R_0 B + \rho_{yx}^A$ [41,45]. A unusual loop-shaped Hall effect (LHE), a hysteresis produced during the upward and downward scan of fields, is also verified in our sample [Supplementary Fig. 2(c)], as reported in previous studies[41,45]. The anomalous Hall resistivity at different temperatures is plotted in Supplementary Fig. 3(a). The anomalous Hall conductivity [$\sigma_{xy}^A = \rho_{yx}^A/({\rho_{yx}^A}^2 + {\rho_{xx}}^2)$] and anomalous Hall angle [$\mathrm{AHA} \equiv \arctan(\sigma_{xy}^A/\sigma_{xx}) \sim \sigma_{xy}^A/\sigma_{xx}$, $\sigma_{xx} = \rho_{xx}/({\rho_{xx}}^2 + {\rho_{yx}}^2)$ shown in Supplementary Fig. 2(d)] are also calculated, as shown in Fig. 1(f). The AHE (AHA) arises below ~ 100 K, and then ascends with temperature approaching the regime of magnetic fluctuations. When the system enters into the ferromagnetic state, the anomalous $\sigma_{xy}^A$ and AHA not vary much.

According to the Mott relation, the thermoelectric signals are proportional to the derivative of the conductivities at $E_F$ [10], which also applies to the anomalous Hall conductivity and anomalous Nernst conductivity[10,57]. Therefore, thermoelectric transport is exquisitely sensitive to the band structure and the anomalous contributions near $E_F$. To further address the anomalous transverse transport in CeAlSi, we performed thermoelectrical transport measurements. Figure 1(g) and Supplementary Fig. 3(b) show the Nernst and magneto-Seebeck signals at selected temperatures, respectively. For Nernst signal, a spinelike profile near zero field can be seen, which maybe comes from conventional contributions as reported in $Cd_3As_2$ [58]. However, for CeAlSi, the spinelike feature only appears when system enters the ferromagnetic state, implying magnetism may adjust the band structure of CeAlSi. The field-independent anomalous components in Nernst are evident at high fields and low temperatures. To further analyze the ANE, an empirical approach is adopted[58]:

$$S_{xy} = S_{xy}^N + S_{xy}^A, \tag{1}$$

$$S_{xy}^N = S_0^N \frac{\mu B}{1+(\mu B)^2}, \tag{2}$$

$$S_{xy}^A = \Delta S_{xy}^A \tanh(B/B_0), \tag{3}$$

Here, $S_{xy}^N$ and $S_{xy}^A$ represent conventional and anomalous contributions, respectively. $S_0^N$, $\Delta S_{xy}^A$, $\mu$, and $B_0$ denote the amplitude of the conventional semiclassical contribution, the amplitude of the anomalous contribution, carrier mobility, and the saturation field above which a plateau appears. From the fit, the amplitude of the anomalous Nernst signal ($|\Delta S_{xy}^A|/T$) is extracted for low temperatures, as shown in the inset of Fig. 1(g). Upon decreasing temperature below ~ 31 K, the ANE appears and

attains a plateau below 15.3 K. There is an abrupt enhancement of the anomalous Nernst angle [ANA ≡ arctan $(\Delta S_{xy}^{A}/S_{xx}) \sim \Delta S_{xy}^{A}/S_{xx}$] when the system enters into the regime of magnetic fluctuations, and then the ANA reaches ~ 9.5% at 7.8 K. The anomalous Nernst conductivity $-\Delta\alpha_{xy}$ is also calculated, which shows similar behavior, as displayed in Supplementary Fig. 3(e). We obtain the contour plots of $\sigma_{xy}^{A}$, $-\Delta\alpha_{xy}$ and $dM/dH$ in Fig. 1(h). As mentioned above, magnetization displays a nonlinear dependence at low temperatures, and this is more evident in the $dM/dH$ plot. Above 15 K, the magnetization is linear, while $\sigma_{xy}^{A}$ and $-\Delta\alpha_{xy}$ already arise, which means that the AHE and ANE do not scale with the magnetization, and they root in topology rather than magnetism. When the system is in the vicinity of the temperature where magnetic fluctuations start to play a role, $\sigma_{xy}^{A}$ and $-\Delta\alpha_{xy}$ are significantly enhanced, implying magnetism interacts with topology, and the interplay between them facilitates the anomalous magneto-transverse transport in CeAlSi. In MTMs, the coupling between magnetic configuration and external magnetic field could produce various intermediate magnetic or topological states, and hence the variation of AHE may come from these states[23–28]. However, for CeAlSi, it was proposed that the angle between the noncollinear spins does not change with applied magnetic field up to 8 T[41]. Therefore, the enhancement of anomalous transverse transport in CeAlSi is supposed to arise from the shift of the positions of Weyl nodes rather than intermediate states.

**Effect of magnetism on the electronic structure of CeAlSi revealed by ARPES**

In order to directly uncover the intricate effect of ferromagnetism on the electronic properties of CeAlSi, we now conduct high-resolution ARPES measurements. In Fig. 2(a), the *f*-electron behavior contributed by Ce is illustrated by the resonant ARPES measurements at the *N* edge of the Ce element. The resonant enhancement at ~ − 0.30 eV, corresponding to the Ce $4f^{1}{}_{7/2}$ final state[59], is revealed at the Ce 4*d* → 4*f* resonant photon energy of 121 eV. The core-level photoemission spectra in Fig. 2(b) show the characteristic peaks of Al-2*p* and Si-2*p* orbitals, where the coexistence of main peaks and shoulders is similar to the case in PrAlSi and SmAlSi[60]. Figure 2(c) presents the bulk and (001)-projected surface BZs with high-symmetry points of the *R*Al*Pn* family. It has been found that the ARPES intensity suffers from the large $k_z$-broadening effect in this series of compounds reflected in the observation of similar band structure in a wide vacuum-ultraviolet photon energy range[38,42,60]. The ARPES spectra would reflect the electronic states integrated over a certain $k_z$ region of the bulk BZ, therefore we use the projected 2D BZ ($\bar{\Gamma}$, $\bar{X}$, $\bar{M}$) hereafter. As shown in Fig. 2(d), the overall Fermi surface (FS) topology of CeAlSi shares many similarities with those of the magnetic *R*Al*Pn* compounds like PrAlSi[60], SmAlSi[39,60], and PrAlGe[38], including the inner ($\alpha$) and outer squarelike pockets around $\bar{\Gamma}$, the dumbbell-like pockets ($\beta$) around $\bar{X}$, and the ripple-shaped FS contours across the BZ boundaries. By a closer look, we notice that there are band splittings of the $\alpha$ and $\beta$ FSs. The splitting of $\beta$ FS has been previously observed in PrAlSi and SmAlSi[60], while the split $\alpha$ FS has not been reported before in other *R*Al*Pn* materials.

To study whether the band splittings are related to ferromagnetism as well as the magnetic impact on the electronic structures of CeAlSi, we then perform temperature-dependent ARPES measurements on another sample. As shown in Figs. 2(e) and 2(f), compared to the other FS contours, the evolution of the $\alpha$ and $\beta$ FSs crossing over $T_c$ is clearly revealed. In the paramagnetic phase [Fig. 2(e)], the splitting of $\alpha$ vanishes and the momentum splitting scale of $\beta$ increases. In Figs. 2(g) and 2(h), we present the near-$E_F$ band dispersions across $T_c$ measured along cuts #a and #b [illustrated in Fig. 2(d)], respectively. It can be seen that the ferromagnetism has sizeable effect on the band structure of CeAlSi, like causing the splitting of $\alpha$ band [Fig. 2(g)] and reducing the splitting of $\beta$ band at $E_F$ [Fig. 2(h)]. Besides, it is also evidenced that the position of $\beta$ band with respect to $E_F$ has also been adjusted via magnetism. Such an evolution of $\beta$ band dispersion indicates that magnetism tune the bulk band structure of CeAlSi as well as the positions of Weyl nodes. The behavior of $\alpha$ is compatible with the traditional Zeeman splitting[61], while the evolution of split $\beta$ could have a more complicated origin, further studies considering the correlation effects of 4*f* electrons would be required. The marked temperature evolutions of $\alpha$ and $\beta$ bands in the energy and momentum imply the intertwined magnetism and itinerant electrons. The current observations are in sharp contrast to the temperature-independent electronic structures in ferromagnetic PrAlSi and antiferromagnetic SmAlSi, where the coupling between the spin configuration of 4*f* states and the conduction electrons has been suggested to be negligible[60]. Moreover, based on the previous results of magnetic *R*Al*Pn* materials[38,39,42,60], the $\alpha$ band is of surface origin, while the origin of $\beta$ band is debated because it can be reproduced by either bulk[60] or surface[39] band calculations. With the presence of the large $k_z$-broadening effect, we deduce that the observed $\beta$ band could be contributed by both the bulk and surface features. This might be a cause of its unusual magnetic behavior. Therefore, the revealed strong interplay between the magnetism and the itinerant electrons both in the bulk and on the surface demonstrates that, among the several *R*Al*Pn* compounds established thus far, CeAlSi could be the most promising platform that hosting the magnetic tunability of bulk Weyl nodes and surface Fermi arcs.

**Pressure evolution of anomalous Hall effect and pressure-induced phase transitions in CeAlSi**

Figure 3(a) displays the resistivity profiles at various pressures. Under pressure, $T_C$ increases monotonically with pressure up to 13.2 GPa [Figs. 3(a) and 3(b)], beyond which it cannot be resolved. The pressure evolution of $T_C$ is roughly consistent with previous reports[40,45]. Above 15.9 GPa, the resistivity shows metallic behavior. Figure 3(c) shows the Hall resistivity at 2 K at various pressures. With increasing pressure to 3.2 GPa, the Hall resistivity decreases slightly, followed by a slight enhancement at 5.6 GPa. Surprisingly, when the pressure reaches 8.5 GPa, the slope of the Hall resistivity changes sign abruptly, indicating a pressure-induced Lifshitz transition (discuss later). We obtain the anomalous Hall resistivity by subtracting the ordinary contribution. Upon increasing pressure, $\rho_{yx}^{A}$ initially decreases, which is consistent with the previous study[45]. However, beyond 1.8 GPa, $\rho_{yx}^{A}$ increases, reaching a maximum at 5.6 GPa. At 8.5 GPa, a sign change from positive to negative accompanied by a slight reduction in $\rho_{yx}^{A}$ implies that the dominant carriers change from hole to electron. Upon further compression, $\rho_{yx}^{A}$ decreases monotonically and then cannot be resolved above 32.5 GPa. The pressure-dependent AHA and the absolute value of anomalous Hall conductivity $|\sigma_{xy}^{A}|$ are also calculated, as plotted in Fig. 3(g), which have a similar evolution as $\rho_{yx}^{A}$ in Fig. 3(d). The anomalous Hall angles are ~ 9.0% and ~ 9.7% for 0.6 GPa and 5.6 GPa, respectively. To further shed light on the intrinsic AHE for pressurized CeAlSi, the anomalous Hall conductivity as a function of the longitudinal conductivity is summarized in Supplementary Fig. 5. For the intrinsic AHE, the anomalous Hall conductivity is independent of the longitudinal conductivity ($\left|\sigma_{xy}^{A}\right| vs. \sigma_{xx}$ ~ constant) [10,62]. Clearly, the data adhere to the universal law both at high and ambient pressures, verifying the intrinsic nature of the AHE in CeAlSi.

The contour profiles of the derivative of the normalized resistivity with respect to temperature, the pressure evolutions of the Hall coefficient ($R_H$), and the resistivity at 2 K are plotted in Figs. 3(e-f). As may be seen, the pressure-induced Lifshitz transition seems to correspond to the evolution of magnetism. To further shed light on the transition, we calculated the magnetic moments under pressure via DFT calculations, yielding 0.8364$\mu_B$, 0.9657$\mu_B$, 0.6811$\mu_B$, and 0.00256$\mu_B$ for 0, 10, 20, and 40 GPa, respectively, which is overall consistent with the experimental data. Thus, the enhancement of $T_C$ under low pressure derives from the pressure-driven enhancement of magnetic moments. Under higher pressure, the magnetic moments decrease gradually, and then disappear, leading to a magnetic phase transition from the ferromagnetic to a paramagnetic state. Since we demonstrate that magnetism has a strong impact on the evolution of band structure at ambient pressure, thus the pressure evolution of magnetic moments provides a strong hint that the Lifshitz transition arises from the coupling between the electronic band structure and magnetic configurations.

To obtain more information about the pressure-induced phase transitions, we investigate the pressure evolution of the crystal structures and electronic band structures. Figure 4(a) displays the high-pressure XRD profiles. Under pressure, the crystal structure with the space group of $I4_1md$ persists up to 39.3 GPa. Upon further compression, a new diffraction peak situated at ~ 10.9° arises, indicative of a structural phase transition. The emerged high-pressure phase coexists with the $I4_1md$ phase up ~ 60 GPa. Figure 4(c) shows the pressure evolution of lattice constants with respect to 1 GPa [Fig. 4(b)] extracted from Rietveld refinements. The ratio of $a/c$ is plotted in the lower panel of Fig. 4(c). As can be seen, in addition to the structural phase transition, there are two anomalies at ~ 10 GPa and ~ 20 GPa, which correspond with the pressures where the Lifshitz transition and the transition from the magnetic state to a paramagnetic state in resistivity appear, respectively. The band structures at several selected pressures are calculated, which remain overall unchanged (Supplementary Fig. 6), except that the hole pockets along the Γ–X line become smaller with pressure and then transform to electron pockets at ~ 10 GPa [Figs. 4(d-i)], which confirms the pressure-induced Lifshitz transition in CeAlSi. This also implies that the pockets along the Γ–X line dominate the transport behavior (the Hall coefficient under pressure changes from positive to negative) in CeAlSi. Since there is no distinct anomaly in the calculated band structures for 10 GPa and 20 GPa, the structural anomalies probably arise from magnetostriction/magnetoelastic effects that are altered by pressure[45,47]. At 0 GPa, the Weyl nodes along the Γ–X line are located 74 meV above $E_F$, whereas they shift to −57 meV and −78 meV below $E_F$ for 10 GPa and 20 GPa, respectively. This indicates that pressure tunes the crystal structure of CeAlSi, which consequently has an effect on the evolution of the Weyl nodes as well as in turn AHE. Based on our calculations, we found that pressure does not change the classification of Weyl nodes, but just shifts the positions of Weyl nodes as observed in magnetism scenario (Supplementary Table 1).

## Discussion

Previous high-pressure studies on CeAlSi revealed a monotonic decline of AHE and LHE with pressure up to 2.7 GPa, while the negligible pressure effect on the magnetic structure and electronic band structure implies the importance of the nontrivial domain walls for the anomalous transport behavior of CeAlSi[43,45]. Under pressure, the anomalous Hall resistivity of CeAlSi is significantly enhanced compared with that at ambient pressure. Considering that the dimensions of the sample (Supplementary Fig. 1) we used are comparable to the size of one single magnetic domain[41,44], the magnetic texture as well as the topological properties can be affected by the domain-wall landscapes or other effects, for example, magnetostriction/magnetoelastic effects[43]. As a consequence, in addition to the tuning of the positions of Weyl nodes via pressure, the contributions to AHE from the landscapes of domain walls need to be elaborated further. Nevertheless, the pressure evolution of AHE together with band calculations provide compelling evidence for the tunability of positions of Weyl nodes via pressure.

The local 4*f*-moments of $Ce^{3+}$ in CeAlSi interact within the lattice, leading to a noncollinear ferromagnetic ordering[41]. ARPES experiments suggest that Ce 4*f* electrons could play a role in CeAlSi, although the band deriving from Ce 4*f* electrons is ~ 0.3 eV below $E_F$ [42]. Supplementary Fig. 8 shows the semilogarithmic plots of the resistivity above 13.2 GPa. The resistivity at 15.9, 17.9 and 19.8 GPa decreases to a minimum, and then shows a $-lnT$ dependence, characteristic of Kondo systems[63]. Below ~ 5 K, the divergence from logarithmic increase suggests the spin-compensated state or the Ruderman-Kittel-Kasuya-Yosida (RKKY) interactions between magnetic impurities may play a role[64,65]. The pressure evolution of magnetic moments and AHE indicates that pressure may also tune the electron correlation effect in CeAlSi. Therefore, the amplified quantum fluctuations in CeAlSi may be recognized as the origin of novel topological states of matter and various quantum phase transitions[66].

In summary, we systematically studied the band structure and topological properties of the ferromagnetic Weyl semimetal CeAlSi through anomalous magneto-transverse transport, ARPES, and band calculations, demonstrating that the positions of Weyl nodes can be tuned via magnetism and pressure. At ambient pressure, the enhancement of AHE and ANE across $T_C$ stems from the shift of the positions of Weyl nodes. The essential role of magnetism in tuning both the bulk and surface band structure of CeAlSi is demonstrated by our ARPES experiments, distinguishing CeAlSi, which has a novel possibility for controlling the Weyl node positions by magnetism, from other magnetic *R*Al*Pn* siblings established thus far. Under pressure, multiple phase transitions are discovered. High-pressure band calculations reveal that pressure could shift the positions of Weyl nodes, which agrees with the transverse transport measurements. Additionally, the domain-wall landscapes, magnetostriction/magnetoelastic effects, or electron correlation effect may also play a significant role in the transport behavior under pressure. These results suggest that ferromagnetic CeAlSi could serve as a fertile and tunable platform to explore novel topological states, and the interplay among magnetism, topology, and electron correlations.

During the preparation of this manuscript, we noticed that the anomalous Nernst effect with different results from ours in CeAlSi has been reported by other workers[67].

## Methods

For the growth of CeAlSi single crystals, a self-flux method was adopted, as described in the literature[32]. The as-grown single crystals were characterized by x-ray diffraction (XRD) measurements [Supplementary Fig. 2]. The details of sample preparation for electrical and thermoelectrical transport measurements, high-pressure electrical transport and XRD measurements, ARPES measurements, and first-principles calculations can be found in the Supplementary Information.

## Data availability

The data that support the findings of this study are available from the corresponding authors upon reasonable request.

## References

1. Yan, B. & Zhang, S.-C. Topological materials. *Rep. Prog. Phys.* **75**, 096501 (2012).

2. Yan, B. & Felser, C. Topological materials: Weyl semimetals. *Annu. Rev. Condens. Matter Phys.* **8**, 337–354 (2017).

3. Jia, S., Xu, S.-Y. & Hasan, M. Z. Weyl semimetals, Fermi arcs and chiral anomalies. *Nat. Mater.* **15**, 1140–1144 (2016).

4. Lv, B. Q., Qian, T. & Ding, H. Experimental perspective on three-dimensional topological semimetals. *Rev. Mod. Phys.* **93**, 025002 (2021).

5. Lv, B., Qian, T. & Ding, H. Angle-resolved photoemission spectroscopy and its application to topological materials. *Nat. Rev. Phys.* **1**, 609–626 (2019).

6. Bernevig, B. A., Felser, C. & Beidenkopf, H. Progress and prospects in magnetic topological materials. *Nature* **603**, 41–51 (2022).

7. Xu, Y. *et al.* High-throughput calculations of magnetic topological materials. *Nature* **586**, 702–707 (2020).

8. Watanabe, H., Po, H. C. & Vishwanath, A. Structure and topology of band structures in the 1651 magnetic space groups. *Sci. Adv.* **4**, eaat8685 (2018).

9. Zou, J., He, Z. & Xu, G. The study of magnetic topological semimetals by first principles calculations. *npj Comput. Mater.* **5**, 96 (2019).

10. Nagaosa, N., Sinova, J., Onoda, S., MacDonald, A. H. & Ong, N. P. Anomalous Hall effect. *Rev. Mod. Phys.* **82**, 1539–1592 (2010).

11. Lancaster, T. Skyrmions in magnetic materials. *Contemp. Phys.* **60**, 246–261 (2019).
12. Qin, Q. *et al.* Emergence of topological Hall effect in a $SrRuO_3$ single layer. *Adv. Mater.* **31**, 1807008 (2019).
13. Vistoli, L. *et al.* Giant topological Hall effect in correlated oxide thin films. *Nat. Phys.* **15**, 67–72 (2019).
14. Kanazawa, N. *et al.* Large topological Hall effect in a short-period Helimagnet MnGe. *Phys. Rev. Lett.* **106**, 156603 (2011).
15. Neubauer, A. *et al.* Topological Hall effect in the A phase of MnSi. *Phys. Rev. Lett.* **102**, 186602 (2009).
16. Kurumaji, T. *et al.* Skyrmion lattice with a giant topological Hall effect in a frustrated triangular-lattice magnet. *Science* **365**, 914–918 (2019).
17. Xu, Y. *et al.* Unconventional transverse transport above and below the magnetic transition temperature in Weyl semimetal $EuCd_2As_2$. *Phys. Rev. Lett.* **126**, 076602 (2021).
18. Tokura, Y. & Kanazawa, N. Magnetic skyrmion materials. *Chem. Rev.* **121**, 2857–2897 (2021).
19. Kanazawa, N. *et al.* Critical phenomena of emergent magnetic monopoles in a chiral magnet. *Nat. Commun.* **7**, 11622 (2016).
20. Fujishiro, Y. *et al.* Topological transitions among skyrmion- and hedgehog-lattice states in cubic chiral magnets. *Nat. Commun.* **10**, 1059 (2019).
21. Puphal, P. *et al.* Topological magnetic phase in the candidate Weyl semimetal CeAlGe. *Phys. Rev. Lett.* **124**, 017202 (2020).
22. Göbel, B., Akosa, C. A., Tatara, G. & Mertig, I. Topological Hall signatures of magnetic hopfions. *Phys. Rev. Res.* **2**, 013315 (2020).
23. Lee, S. H. *et al.* Evidence for a magnetic-field-induced ideal type-II Weyl state in antiferromagnetic topological insulator $Mn(Bi_{1-x}Sb_x)_2Te_4$. *Phys. Rev. X* **11**, 031032 (2021).
24. Li, J. *et al.* Intrinsic magnetic topological insulators in van der Waals layered $MnBi_2Te_4$-family materials. *Sci. Adv.* **5**, eaaw5685 (2019).
25. Su, H. *et al.* Magnetic exchange induced Weyl state in a semimetal $EuCd_2Sb_2$. *APL Mater.* **8**, 011109 (2020).
26. Ma, J.-Z. *et al.* Spin fluctuation induced Weyl semimetal state in the paramagnetic phase of $EuCd_2As_2$. *Sci. Adv.* **5**, eaaw4718 (2019).
27. Cheng, E. J. *et al.* Magnetism-induced topological transition in $EuAs_3$. *Nat. Commun.* **12**, 6970 (2021).
28. Yuan, J. *et al.* Magnetization tunable Weyl states in $EuB_6$. *Phys. Rev. B* **106**, 054411 (2022).

29. Gaudet, J. *et al.* Weyl-mediated helical magnetism in NdAlSi. *Nat. Mater.* **20**, 1650–1656 (2021).
30. Chang, G. *et al.* Magnetic and noncentrosymmetric Weyl fermion semimetals in the *R*AlGe family of compounds (*R* = rare earth). *Phys. Rev. B* **97**, 041104 (2018).
31. Xu, S.-Y. *et al.* Discovery of Lorentz-violating type II Weyl fermions in LaAlGe. *Sci. Adv.* **3**, e1603266 (2017).
32. Su, H. *et al.* Multiple Weyl fermions in the noncentrosymmetric semimetal LaAlSi. *Phys. Rev. B* **103**, 165128 (2021).
33. Zhao, J. *et al.* Field-induced tricritical phenomenon and magnetic structures in magnetic Weyl semimetal candidate NdAlGe. *New J. Phys.* **24**, 013010 (2022).
34. Wang, J.-F. *et al.* NdAlSi: A magnetic Weyl semimetal candidate with rich magnetic phases and atypical transport properties. *Phys. Rev. B* **105**, 144435 (2022).
35. Yang, R. *et al.* Charge dynamics of a noncentrosymmetric magnetic Weyl semimetal. *npj Quantum Mater.* **7**, 101 (2022).
36. Destraz, D. *et al.* Magnetism and anomalous transport in the Weyl semimetal PrAlGe: possible route to axial gauge fields. *npj Quantum Mater.* **5**, 5 (2020).
37. Lyu, M. *et al.* Nonsaturating magnetoresistance, anomalous Hall effect, and magnetic quantum oscillations in the ferromagnetic semimetal PrAlSi. *Phys. Rev. B* **102**, 085143 (2020).
38. Sanchez, D. S. *et al.* Observation of Weyl fermions in a magnetic non-centrosymmetric crystal. *Nat. Commun.* **11**, 3356 (2020).
39. Zhang, Y. *et al.* Kramers nodal lines and Weyl fermions in SmAlSi. Preprint at http://arxiv.org/abs/2210.13538 (2022).
40. Cao, W. *et al.* Quantum oscillations in noncentrosymmetric Weyl semimetal SmAlSi. *Chin. Phys. Lett.* **39**, 047501 (2022).
41. Yang, H.-Y. *et al.* Noncollinear ferromagnetic Weyl semimetal with anisotropic anomalous Hall effect. *Phys. Rev. B* **103**, 115143 (2021).
42. Sakhya, A. P. *et al.* Observation of Fermi arcs and Weyl nodes in a non-centrosymmetric magnetic Weyl semimetal. Preprint at http://arxiv.org/abs/2203.05440 (2022).
43. Xu, B. *et al.* Picoscale magnetoelasticity governs heterogeneous magnetic domains in a noncentrosymmetric ferromagnetic Weyl semimetal. *Adv. Quantum Technol.* **4**, 2000101 (2021).
44. Sun, Y. *et al.* Mapping domain-wall topology in the magnetic Weyl semimetal CeAlSi. *Phys. Rev. B* **104**, 235119 (2022).
45. Piva, M. M. *et al.* Topological features in the ferromagnetic Weyl semimetal CeAlSi: Role of domain walls. *Phys. Rev. Res.* **5**, 013068 (2023).

46. Hodovanets, H. *et al.* Anomalous symmetry breaking in the Weyl semimetal CeAlGe. *Phys. Rev. B* **106**, 235102 (2022).

47. He, X. *et al.* Pressure tuning domain-wall chirality in noncentrosymmetric magnetic Weyl semimetal CeAlGe. Preprint at http://arxiv.org/abs/2207.08442 (2022).

48. Xu, L. *et al.* Shubnikov–de Haas oscillations and nontrivial topological states in Weyl semimetal candidate SmAlSi. *J. Phys. Condens. Matter* **34**, 485701 (2022).

49. Suzuki, T. *et al.* Singular angular magnetoresistance in a magnetic nodal semimetal. *Science* **365**, 377–381 (2019).

50. Ng, T. *et al.* Origin and enhancement of the spin Hall angle in the Weyl semimetals LaAlSi and LaAlGe. *Phys. Rev. B* **104**, 014412 (2021).

51. MacNeill, D. *et al.* Control of spin–orbit torques through crystal symmetry in $WTe_2$/ferromagnet bilayers. *Nat. Phys.* **13**, 300–305 (2017).

52. Zhou, J., Qiao, J., Bournel, A. & Zhao, W. Intrinsic spin Hall conductivity of the semimetals $MoTe_2$ and $WTe_2$. *Phys. Rev. B* **99**, 060408 (2019).

53. Shi, S. *et al.* All-electric magnetization switching and Dzyaloshinskii–Moriya interaction in $WTe_2$/ferromagnet heterostructures. *Nat. Nanotechnol.* **14**, 945–949 (2019).

54. Stiehl, G. M. *et al.* Layer-dependent spin-orbit torques generated by the centrosymmetric transition metal dichalcogenide $\beta$−$MoTe_2$. *Phys. Rev. B* **100**, 184402 (2019).

55. Cao, W. *et al.* Pressure-induced superconductivity in the noncentrosymmetric Weyl semimetals LaAl*X* (*X* = Si, Ge). *Phys. Rev. B* **105**, 174502 (2022).

56. Gaudet, X. Y. J. *et al.* Topological spiral magnetism in the Weyl semimetal SmAlSi. Preprint at http://arxiv.org/abs/2206.05121 (2022).

57. Fu, C. G., Sun, Y. & Felser, C. Topological thermoelectrics. *APL Mater.* **8**, 040913 (2020).

58. Liang, T. *et al.* Anomalous Nernst effect in the Dirac semimetal $Cd_3As_2$. *Phys. Rev. Lett.* **118**, 136601 (2017).

59. Allen, J. W. The Kondo resonance in electron spectroscopy. *J. Phys. Soc. Jpn* **74**, 34–48 (2005).

60. Lou, R. *et al*. Signature of weakly coupled *f* electrons and conduction electrons in magnetic Weyl semimetal candidates PrAlSi and SmAlSi. *Phys. Rev. B* **107**, 035158 (2023).

61. Petrovykh, D. Y. *et al*. Spin-dependent band structure, Fermi surface, and carrier lifetime of permalloy. *Appl. Phys. Lett.* **73**, 3459–3461 (1998).

62. Miyasato, T. *et al.* Crossover behavior of the anomalous Hall effect and anomalous Nernst effect in itinerant ferromagnets. *Phys. Rev. Lett.* **99**, 086602 (2007).

63. Kondo, J. Resistance minimum in dilute magnetic alloys. *Prog. Theo. Phys.* **32**, 37–49 (1964).
64. Liang, V. K. C. & Tsuei, C. C. Kondo effect in an amorphous $Ni_{41}Pd_{41}B_{18}$ alloy containing Cr. *Phys. Rev. B* **7**, 3215 (1973).
65. Ding, X. X. *et al*., Crossover from Kondo to Fermi-liquid behavior induced by high magnetic field in 1*T*-$VTe_2$ single crystals. *Phys. Rev. B* **103**, 125115 (2021).
66. Paschen, S. & Si, Q. Quantum phases driven by strong correlations. *Nat. Rev. Phys*. **3**, 9–26 (2021).
67. Alam, M. S. *et al*., Sign change of the anomalous Hall effect and the anomalous Nernst effect in Weyl semimetal CeAlSi. Preprint at http://arxiv.org/abs/2210.09764 (2022).

## Acknowledgments

This work is supported by the Deutsche Forschungsgemeinschaft (DFG) through the project C03, C04 and C07 of the Collaborative Research Center SFB 1143 (project-ID 247310070), the Würzburg-Dresden Cluster of Excellence on Complexity and Topology in Quantum Matter—ct.qmat (EXC 2147, Project No. 390858490). E.J.C. acknowledges the financial support from the Alexander von Humboldt Foundation. W.G.Y. acknowledges the National Natural Science Foundation of China (Grant No. U1930401). S.Y.L. acknowledges the National Natural Science Foundations of China (Grant No. 12174064), and the Ministry of Science and Technology of China (Grant No. 2022YFA1402203). W.W.Z. is supported by the Shenzhen Peacock Team Plan (KQTD20170809110344233) and the Bureau of Industry and Information Technology of Shenzhen through the Graphene Manufacturing Innovation Center (201901161514). Y.F.G. acknowledges research funds from the State Key Laboratory of Surface Physics and Department of Physics, Fudan University (Grant No. KF2019 06). Y.J.X. was supported by the National Natural Science Foundation of China (Grant No. 12204033).

## Author Contributions

E.J.C. conceived the idea and designed the experiments. E.J.C., J.Y., W.X., and Y.F.G. prepared the single crystals. E.J.C. conducted magnetic susceptibility measurements and electrical transport measurements at ambient pressure, and thermoelectric measurements with the help of M.B.. E.J.C. and L.M.Y. were responsible for electrical transport experiments under pressure. L.M.Y. and W.G.Y. conducted the high-pressure XRD measurements and analysis. X.B.S. and W.W.Z. performed the DFT calculations for the pressure evolution of the electronic band structure. R.L. and A.F. conducted ARPES experiments. Y.M.W. helped with the data collection. Y.J.X. performed the DFT calculations for the pressure evolution of the magnetic moments. N.P. and D.C.P. helped with the orientation of the single crystals for thermoelectric measurements. E.J.C., L.M.Y., X.B.S., R.L. and Y.X. analyzed the data. E.J.C., R.L., W.G.Y. and B.B. supervised the project. E.J.C., L.M.Y., X.B.S. and R.L. contributed equally to this work. E.J.C. wrote the paper with input from all coauthors.

## Competing interests

The authors declare no competing interests.

## Additional Information

**Supplementary information** is available for this paper at the URL inserted when published.

**Correspondence** and requests for materials should be addressed to E.J.C. (erjian_cheng@163.com), R.L. (lourui@lzu.edu.cn), W.G.Y. (yangwg@hpstar.ac.cn), or B.B. (B.Buechner@ifw-dresden.de).

## Figure captions

**Figure 1 | Anomalous Hall effect (AHE) and anomalous Nernst effect (ANE) in CeAlSi. a** The schematic structure of CeAlSi with a noncentrosymmetric structure (space group of $I4_1md$). **b** Band structure and associated Berry curvature. **c** Longitudinal resistivity ($\rho_{xx}$) in zero field. Zero-field-cooling (ZFC) and field-cooling (FC) magnetization as a function of temperature for CeAlSi with the magnetic field applied along the *c* axis. The ferromagnetic transition temperature ($T_C$) is ~ 10.1 K defined according to magnetization. **d** Transverse Hall resistivity ($\rho_{yx}$) different temperatures with the magnetic field applied along the *c* axis. **e** Field dependence of magnetization at various temperatures with the magnetic field applied along the *c* axis. Inset shows the low-field data. **f** Anomalous Hall conductivity ($\sigma_{xy}^{A}$) at various temperatures. Inset displays the anomalous Hall angle (AHA). **g** Nernst signal normalized to the temperature at different temperatures. Inset shows the temperature dependence of the amplitude of anomalous Nernst signal normalized to the temperature ($|S_{xy}^{A}|/T$), and the anomalous Nernst angle (ANA). **h** Contour plots of the $\sigma_{xy}^{A}$, $-\Delta\alpha_{xy}$ (see Supplementary Note 3 for more details) and the derivative of magnetization ($dM/dH$). The background color represents the magnitude of their values.

**Figure 2 | ARPES measurements of CeAlSi. a** Angle-integrated photoemission spectra of CeAlSi with Ce *N* edge on-resonant (121 eV) and off-resonant (116 eV) photons, respectively. **b** Core-level photoemission spectra of CeAlSi recorded at $h\nu$ = 160 eV. **c** Sketches of 3D BZ and (001)-surface BZ for the noncentrosymmetric $I4_1/md$ space group structure. **d** Constant-energy ARPES image of CeAlSi ($h\nu$ = 40 eV, $CR^+$ polarization, $T$ = 1.4 K, sample no. S1) obtained by integrating the photoemission intensity within $E_F \pm 20$ meV. Cuts #a and #b indicate the locations of the band dispersions in **g** and **h**, respectively. **e**, **f** Constant-energy maps at $E_F$ of CeAlSi ($h\nu$ = 40 eV, $CR^+$ polarization, sample no. S2) taken above (13 K) and below (1.5 K) $T_c$, respectively. The red solid curves in **d**-**f** represent the (001)-projected BZs. $a$ (= 4.26 Å) is the in-plane lattice constant of CeAlSi. **g** Second derivative intensity plots of CeAlSi measured along cut #a above and below $T_c$, respectively. The black arrows indicate the splitting of $\alpha$ band below $T_c$. **h** Same as **g** recorded along cut #b. The red solid curves are guides to the eye for the split $\beta$ bands.

**Figure 3 | Pressure-induced phase transitions in CeAlSi. a** Temperature dependence of longitudinal resistivity at different pressures. **b** Low-temperature resistivity normalized to the data at 50 K. With

increasing pressure, the ferromagnetic transition temperature ($T_C$) initially increases. **c** Hall resistivity at various pressures. Above 5.6 GPa, the slope of Hall resistivity changes sign, indicating that the dominant carriers change from holes to electrons. Inset shows the high-pressure data above 15.9 GPa. **d** Anomalous Hall resistivity ($\rho_{yx}^{\mathrm{A}}$) at various pressures. Inset shows the high-pressure data. **e** Contour plot of the derivative of normalized resistivity at different pressures. The background color represents the $d(\rho_{xx}/\rho_{xx}^{50\mathrm{K}})/dT$ value. The pressure evolution of $T_C$ is added. **f** Pressure-dependent Hall coefficient ($R_{\mathrm{H}}$) and the resistivity at 2 K ($\rho_{xx}^{2\mathrm{K}}$). $R_{\mathrm{H}}$ is obtained through linear fits to the high-field data. The shaded area represents the pressure region where $R_{\mathrm{H}}$ changes sign, suggesting the existence of a pressure-induced Lifshitz transition. **g** Pressure dependence of anomalous Hall angle (AHA) and absolute value of anomalous Hall conductivity ($|\sigma_{xy}^{\mathrm{A}}|$).

**Figure 4 | Pressure evolution of the crystal structure and band structure of CeAlSi. a** X-ray diffraction (XRD) pattern of CeAlSi at room temperature up to 60 GPa. The ambient-pressure structure with the space group of $I4_1md$ persists to ~ 39.3 GPa, beyond which a new diffraction peak emerges (marked with a dashed line and asterisk), indicating that a pressure-induced structural phase transition occurs. 0′ represents that the pressure inside the sample chamber is released to zero, indicating that the emerging new structural phase is unstable at ambient pressure. **b** The Rietveld refinement of the XRD pattern at 1.0 GPa. The refined value is $R_{\mathrm{P}}$ = 2.23% with weighted profile $R_{\mathrm{WP}}$ = 1.60%. The upper panel in (**c**) shows the pressure-dependent normalized parameters $a/a_0$, $c/c_0$ and $V/V_0$ extracted from powder diffraction refinements. The lower panel in (**c**) shows the pressure evolution of the $a/c$ ratio. **d-f** Band structures of CeAlSi along the Γ-W-X line for 0 GPa, 10 GPa, and 20 GPa, respectively. **g-i** Calculated 3-dimensional (3D) Fermi surfaces for 0 GPa, 10 GPa, and 20 GPa, respectively. The violet and dark yellow color represent electron pockets and hole pockets, respectively. At 10 GPa, pressure drives hole pockets (the red dashed circle as marked in **g**) into electron pockets, demonstrating the pressure-induced Lifshitz transition observed in Hall resistivity under pressure.

# Figure 1

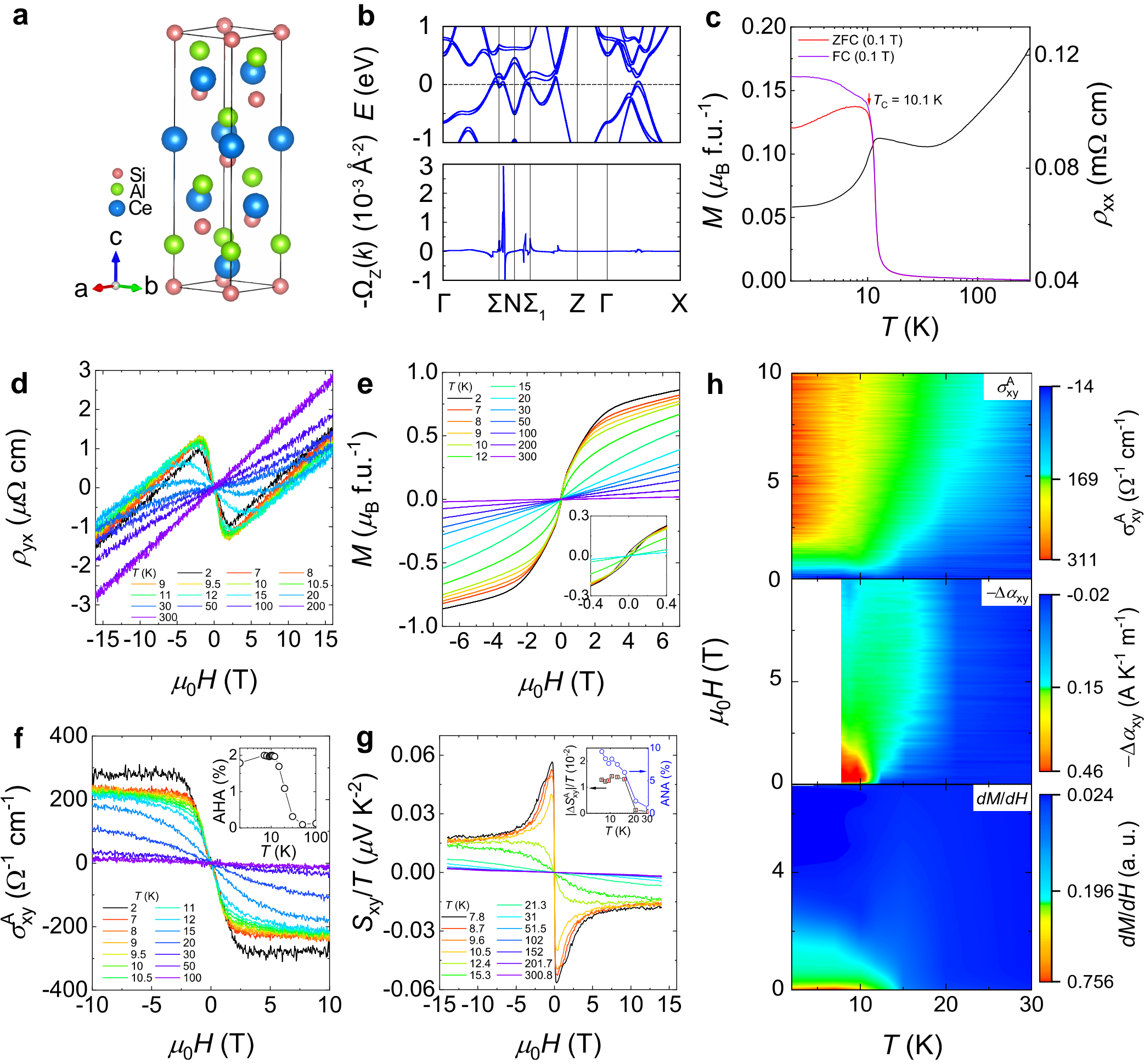

Figure 2

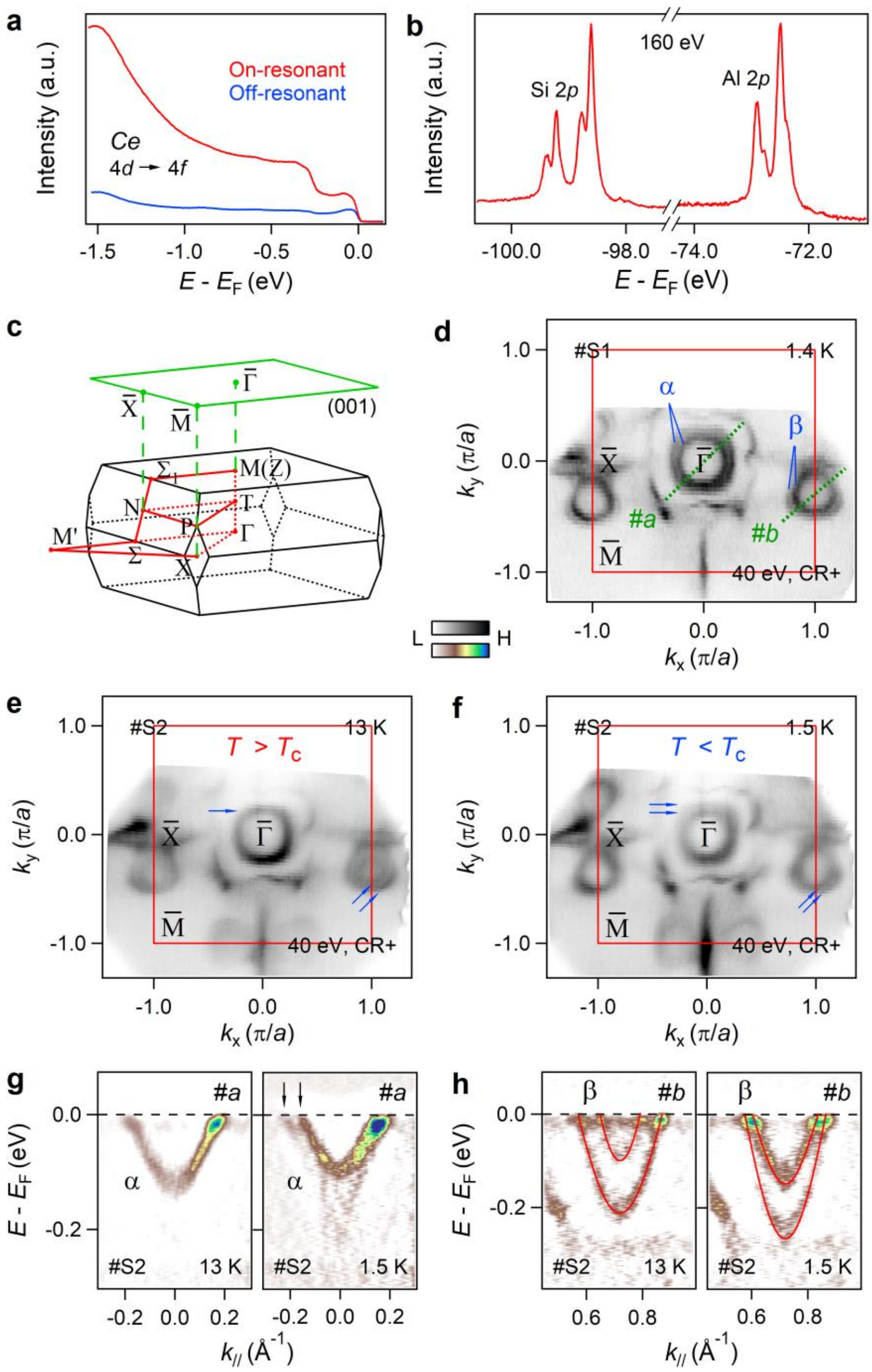

Figure 3

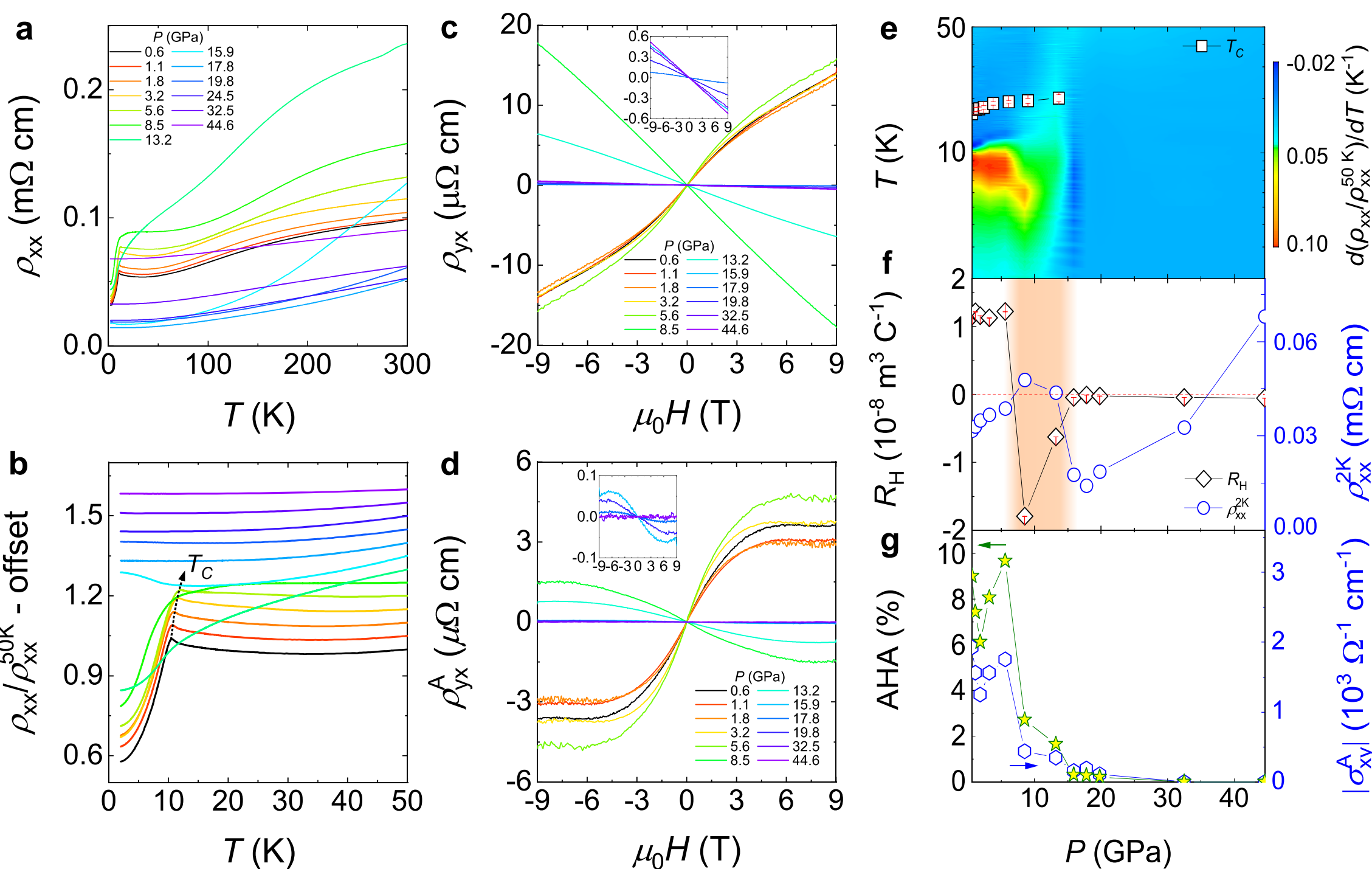

# Figure 4

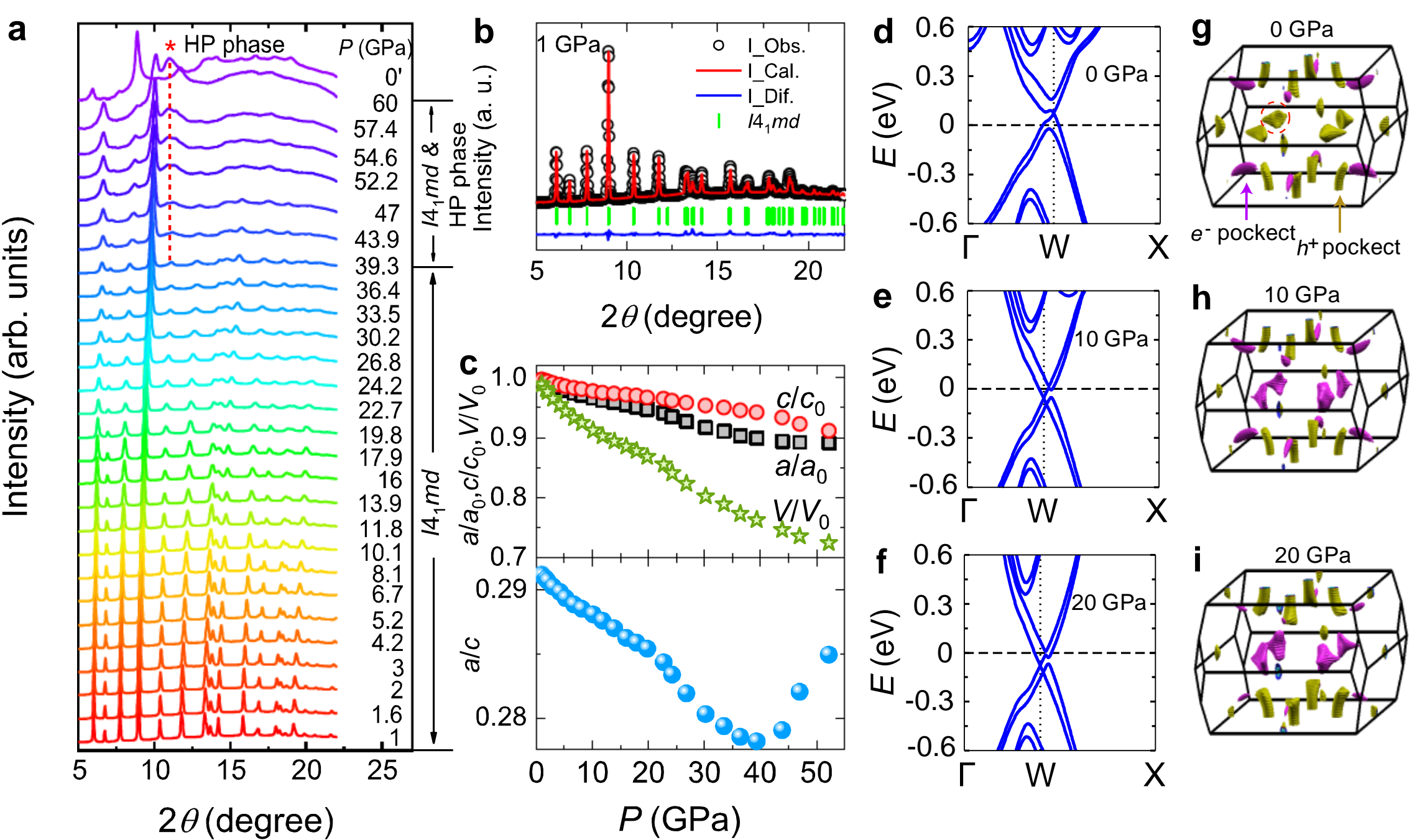